\documentclass[twocolumn,showpacs,amssymb,preprintnumbers,prl,floatfix]{revtex4-1}
\usepackage{epsfig,dcolumn}
\usepackage{bm,color}
\usepackage{graphicx,bm}
\newcommand{\be}{\begin{equation}}
\newcommand{\ee}{\end{equation}}
\newcommand{\ba}{\begin{eqnarray*}}
\newcommand{\ea}{\end{eqnarray*}}

\begin{document}
\title{Broken mirror symmetry in $^{36}$S and $^{36}$Ca}

\author{J.J. Valiente-Dob\'on$^{1}$, A.~Poves$^{2}$,  A. Gadea$^{3}$,  B. Fern\'andez-Dom\'{\i}nguez$^{4}$}

\affiliation{$^1$Istituto Nazionale di Fisica Nucleare, Laboratori Nazionali di Legnaro, Legnaro, Italy}
\affiliation{$^2$Departamento de F\'isica Te\'orica and IFT-UAM/CSIC, Universidad Aut\'onoma de Madrid,  E-2804 Madrid, Spain}
\affiliation{	$^3$Instituto de F\'isica Corpuscular, CSIC-Universidad de Valencia, Valencia, Spain}    
\affiliation{$^4$Universidade de Santiago de Compostela, E-15782 Santiago de Compostela, Spain}

\begin{abstract}

  Shape coexistence is an ubiquitous phenomenon in  the neutron-rich nuclei belonging to (or sitting at the shores of) the $N=20$ Island of Inversion (IoI).
  Exact isospin symmetry predicts the same behaviour for their mirrors and the  existence of a proton-rich IoI around $Z=20$, centred
  in the (surely unbound) nucleus $^{32}$Ca. In this article we show that in  $^{36}$Ca  and  $^{36}$S,  Coulomb effects break dramatically the mirror  
   symmetry in the excitation energies,
  due to the different structures of the intruder and normal states. 
  The Mirror Energy Difference (MED)  of their 2$^+$ states is known to be very large at -246 keV.   
  We reproduce this value and predict the first excited  state in  $^{36}$Ca to be a 0$^+$ at 2.7 MeV, 250 keV below   
  the first 2$^+$.  In its mirror $^{36}$S  the 0$^+$  lies  at 55 keV  above  the  2$^+$ measured at 3.291 MeV.  Our
  calculations predict a huge  MED
  of  -720 keV,  that  we  dub   "Colossal" Mirror Energy Difference (CMED).  
  A possible reaction mechanism to access the 0$^+_2$ in $^{36}$Ca will be discussed. 
  In addition, we theoretically address  the  MED's of  the $A=34$ $T=3$ and $A=32$ $T=4$ mirrors.

  \end{abstract}


\date{\today}
\maketitle

 The study of the effects of the Isospin Symmetry Breaking (ISB) terms of the nucleon-nucleon interaction on nuclear properties, 
 particularly the Coulomb repulsion among the protons, has a long-standing
 history, starting with the Nolen and Schiffer anomaly  \cite{Nolen_Schiffer}, which involves the mass difference of a pair of mirror nuclei, and following
 up with their effects in spectroscopic properties like the Mirror Energy Differences (MED) and the Triplet Energy Differences (TED) extracted from the comparison of the
 excitation spectra of the members of an isobaric multiplet \cite{Zuker_prl,Bentley_Lenzi}. These studies have shown that the MED's reflect nicely some
 structural properties of the states in question, such as deformation, alignment, occupancies of particular orbits, etc.  On another register, the study of 
 neutron-rich nuclei near the neutron magic shell closures has lead to  the discovery of the so called Islands of Inversion (IoI's), groups of nuclei which, 
 unexpectedly, have their ground states dominated by intruder configurations, most often of deformed nature. The relevant IoI for  our present
 purpose is  at $N=20$, centred about  $^{31}$Na \cite{1975thibault,Huber:1978co,detraz,GuillemaudMueller,baumann,campi,Poves:1987jg,warburton,heyde,fukunishi}. 
 At or around these IoI's it is very frequent to find states of different shapes coexisting
 in the same nucleus.  Therefore, if isospin symmetry holds (and we know it does to a very large extent) each IoI at the neutron-rich side should
 have a mirror IoI at the proton-rich side. However, only for relatively light nuclei can one hope to reach or even to approach such proton IoI's. For the 
 $Z=20$ isotopes, $^{32}$Ca most likely is experimentally out of reach, perhaps we can reach $^{34}$Ca, and  there is already some information 
 about  $^{36}$Ca in refs.~\cite{Doornenbal:2007cv,Burger:2012gh}.

  The structure and location of  coexisting intruder 0$^+$ states at $N=20$ evolves as we move away from  N=Z. Indeed, the first excited state in  
  $^{40}$Ca at 3.353 MeV is the head of a deformed band of 4p-4h nature. The relevant experimental information about the $A=38$ $T=1$ mirrors
  is gathered in Table~\ref{tab:A38}. The second  0$^+$ and  2$^+$ states are again of intruder nature; neutron 2p-2h in $^{38}$Ar   and 
  proton  2p-2h in $^{38}$Ca. And this is clearly manifested in their MED's which are very large, because the two protons promoted to the $pf$ shell
  suffer less Coulomb repulsion than when occupying the $sd$ shell, that is why the excitation energies of the intruders are reduced in $^{38}$Ca. With this
  anchor we can proceed further into the proton-rich side $A=36$ looking for an enhanced ISB effect associated with shape coexistence.

\begin{table}[h]
\caption{\label{tab:A38} Experimental excitation energies (in MeV) and MED's (in keV) for the mirror nuclei $A=38$ $T=1$.}
\begin{tabular*}{\linewidth}{@{\extracolsep{\fill}}|ccccc|}
\hline  
J$^{\pi}$ & $^{38}$Ca (exp)&  $^{38}$Ar (exp)  & MED (exp) & MED(th)\\ 
\hline
 $0^+_1$  & 0.0 &  0.0     &      &        \\ 
 $2^+_1$  & 2.213 &  2.168 &  +45  & -25\\ 
  $0^+_2$  & 3.084 & 3.378       &  -294   & -340  \\ 
   $2^+_2$  & 3.684 & 3.936       &  -252   & -340 \\ 
\hline    
 \end{tabular*}
\end{table}   

$^{36}$S is stable and extensively studied experimentally.  We list the states of interest in Table~\ref{tab:A36}, the normal 0$^+$ and  2$^+$ states and the intruder 0$^+$.  
For  $^{36}$Ca the only spectroscopic information available is the excitation energy of the first 2$^+$ state~\footnote{An experiment of relativistic Coulomb excitation at  RIKEN, only published in conference proceedings~\cite{iwasa}, claims that \mbox{B(E2;~0$_{gs}^+~\rightarrow~2^+$)~=~71$^{+17}_{-13}$~$e^2fm^4$} in $^{36}$Ca, a value five times larger than the USD prediction \cite{usd}} from refs.~\cite{Doornenbal:2007cv,Burger:2012gh}.
  Note the very large experimental MED for the spherical 2$^+$ state of \mbox{-246(3)~keV}, at variance
  with the situation in the $A=38$ mirrors, where the MED is +45 keV. Indeed this large shift in the $A=36$ mirrors cannot have the same
  origin as the ones found in the intruder states of $A=38$.

   We proceed now with the theoretical description of the $A=36$  T=2 mirrors in the framework of the Shell Model with Configuration
    Interaction \cite{Cau05}. We adopt
   the valence space and the effective interaction (sdpfu-mix) which has been successfully applied in the simultaneous description
    of the $N=20$ and $N=28$ IoI's
   in ref.~\cite{n2028}. We add to the nuclear interaction  the two-body matrix elements of the Coulomb potential computed in an oscillator basis with the
   appropriate oscillator parameter \mbox{$\hbar \omega$= 45 A$^{-1/3}$ - 25 A$^{-2/3}$}.    
   The proton $sd$-shell single particle energies (SPE) could be derived from the experimental 
   spectra of $^{17}$F  and that of the proton $pf$-shell orbits from  the spectrum of 
   $^{41}$Sc. The value of the Coulomb shift of the 1s$_{1/2}$ proton single particle energy relative to the corresponding neutron single particle energy
   would then be 375~keV and that of the 1p$_{3/2}$ and 
   1p$_{1/2}$ proton  orbits  200~keV. However, it is seen experimentally that the MED's in the  mirrors   $^{39}$Ca - $^{39}$K
   and    $^{37}$Ca - $^{37}$Cl are much smaller, 56~keV and 120~keV respectively. Drawing from the findings of ref. \cite{blz}
   which concludes that the 1s$_{1/2}$ orbit has a very large radius when empty at the mean field level, independent of any energy threshold  
   effect,  becoming smaller as it is filled. We take an interpolated value of 300~keV for the shift in A=36. 
   However, these SPE's have unwanted effects in some MED's directly related to the Z=14 gap.
   Hence,   following the analysis of ref~\cite{Doornenbal:2007cv}, we have resorted to a minimal modification of the proton SPE's. Our
   ansatz is the following: the  Z=14 proton gap remains unchanged 
   whereas the Z=16 gap is reduced by 300 keV. This choice of the proton
   SPE's  results in a MED for the  $^{29}$S - $^{29}$Al mirror pair of --46~keV, in reasonable agreement with the experimental 
   value --176(20)~keV from ref. \cite{Reynolds}.   
  The experimental  $Z = 20$ and $N = 20$ shell gaps at  $^{40}$Ca are essentially equal, they differ by just 29 keV,   
  our calculations reproduce nicely this difference, a theoretical value of 27~keV is obtained.
  The MED's of the A=38 mirror pair are also well reproduced, as can be seen in Table \ref{tab:A38}. 
    Since the choice of proton SPE's is irrelevant for this case and the neutron and proton gaps are equal, the large MED's of the 
    intruder states have their origin only on the two-body Coulomb repulsion.

 \begin{table}
\caption{\label{tab:A36} Excitation energies (in MeV) and MED's (in keV). In the column labeled "$A=36$ $T=2$" we list the results of a 
 calculation without the Coulomb interaction.}
\begin{tabular*}{\linewidth}{@{\extracolsep{\fill}}|ccccccc|}
\hline  
J$^{\pi}$ & A=36 T=2 &$^{36}$Ca (exp)& (th) & $^{36}$S (exp) & th & MED (th)\\ 
\hline
 $0^+_1$  & 0.0 &  0.0     &  0.0  &       0.0   & 0.0&   \\ 
 $2^+_1$  & 2.97 & 3.045 &  2.95 &     3.291 & 3.25 & -300 \\ 
  $0^+_2$  &  2.97 &        &  2.70 &     3.346 &  3.42 & -720\\ 
\hline  
\end{tabular*}
\end{table}  

\begin{figure}[h]
\begin{center}
\includegraphics[width=\columnwidth,angle=0]{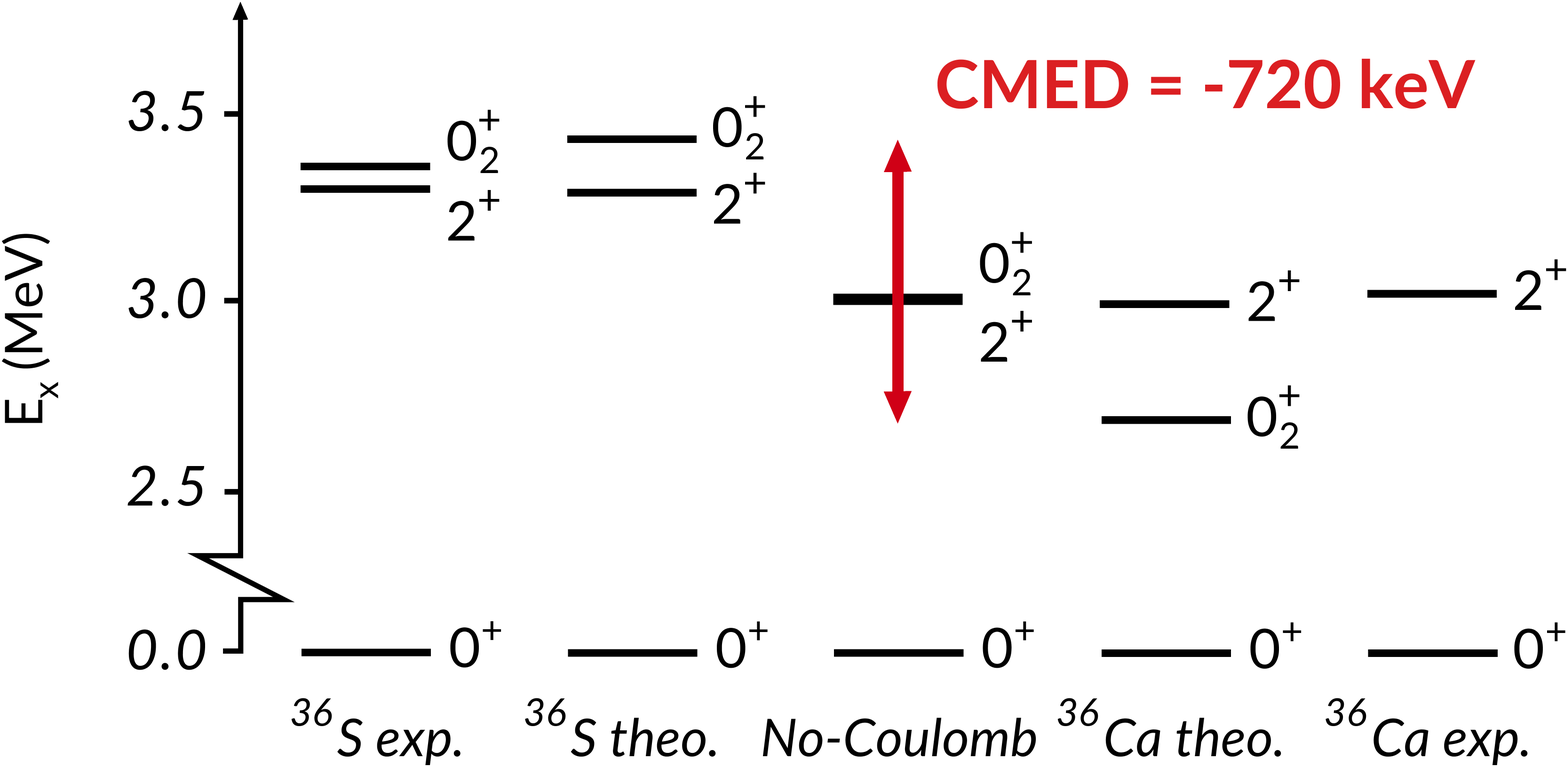}
\end{center}
\caption{(Colour online) Low-lying excited states in the mirror pair $^{36}$Ca - $^{36}$S. The known experimental information is shown together with shell-model calculations using the sdpfu-mix interaction. At the center of the figure, the results without the Coulomb interaction are shown. The main configurations for $^{36}$S are:  0$_{g.s.}^+$  \mbox{$d_{5/2}^6 s_{1/2}^2$} (protons) and $(sd)^{12}$ (neutrons); 2$^+$  \mbox{$d_{5/2}^6 s_{1/2}^1 d_{3/2}^1$} (protons) and $(sd)^{12}$ (neutrons). Instead, for the intruder second 0$^+$ the main configuration is: \mbox{$d_{5/2}^6 s_{1/2}^1 d_{3/2}^1$} (protons) \mbox{$d_{5/2}^6 s_{1/2}^2 d_{3/2}^2 (pf)^2$} (neutrons). For $^{36}$Ca it suffices to exchange the role of protons and neutrons. \label{level}}
\end{figure}

 The results for the mirror pair $^{36}$Ca - $^{36}$S are shown in Table~\ref{tab:A36}. 
 The calculation reproduces the large MED of the 2$^+$, with the same mechanism 
  discussed in Ref.~\cite{Doornenbal:2007cv}.  The origin of this large MED is easily grasped if we 
  compare the spectra of $^{36}$S and  $^{36}$Ca, shown in  Fig.~\ref{level}, with the spectrum obtained
  in the calculation without the Coulomb interaction. Whereas the excitation energy of the 2$^+$ in
  $^{36}$Ca barely moves with respect to the no-Coulomb reference, in $^{36}$S  it goes up  by 280~keV.
  The reason lies in the fact that the proton 1s$_{1/2}$ orbit is more tightly bound than the neutron 1s$_{1/2}$, relative to the corresponding 0d$_{3/2}$-orbits.
  As the configuration of the 2$^+$  is  1s$_{1/2}^{1}$  0d$_{3/2}^1$ the result follows trivially.  But what happens
  for the intruder  0$^+$ state? Let's compare again the two mirrors with the  no-Coulomb case; in $^{36}$S  the  0$^+_2$ 
  excitation energy increases by the same amount as in the 2$^+$ case. And this may seem unexpected because one
  might na\"ively think that its  proton configuration is close to 1s$_{1/2}^{2}$. Which is not the case indeed, because due to the
  deformed nature of the intruder band, the $sd$ shell occupancies approach the pseudo-SU3 limit, being rather  
  close to 1s$_{1/2}^{1}$  0d$_{3/2}^1$. Moving to  $^{36}$Ca, the proton configuration becomes $(sd)^{10}$-$(pf)^2$ which, as
  discussed for the $A=38$ pair, has less Coulomb repulsion than the $(sd)^{12}$  configuration of the  0$^+$ ground state.
  These two shifts of quite different origin add constructively to produce a Colossal Mirror Energy Difference (CMED) of \mbox{-720 keV},
  without advocating energy threshold effects. This is our main prediction.
  As a consequence, the intruder 0$^+$ becomes the first excited state of $^{36}$Ca , decaying by an $E0$ transition to the ground state.
   We do not expect energy threshold effects due to the proximity of the excitation energy of the  0$^+$ intruder to the 2p separation energy, because 
  of the Coulomb barrier which makes the (less bound) 2$^+$ a very narrow state (the one- and two-proton separation energies, S(p) and S(2p)
   are about  2.6~MeV in $^{36}$Ca).
   Our prediction agrees nicely with what can be na\"ively expected from the known experimental MED nearby. Indeed, the experimental MED 
  of the excited  0$^+$  state of the A=38 mirrors gives a hint of the extra contribution to the MED in the case of intruder states, whereas  
  the experimental MED of the 2$^+$ state in the  $^{36}$Ca - $^{36}$S mirror pair, does the same for the contribution to the MED of a configuration
   1s$_{1/2}^{1}$  0d$_{3/2}^1$. 
  Knowing from theory that this is indeed the neutron(proton) configuration in the intruder 0$^+$ of $^{36}$Ca($^{36}$S), one can conclude that 
  both contributions add constructively to produce a MED of about --600~keV. In this discussion we have 
  not adopted any theoretical ansatz for the one body and two body Coulomb effects, we have just made an educated guess drawing  from the 
  available experimental data.

   There are  a few other known cases of  MED's of similar size. However, all of them are  dominated by energy threshold effects, i.e. they
   involve an excited state with  an important  1s$_{1/2}$  content which is well above the  proton separation energy of the proton rich
   mirror.  For instance,  the
  \mbox{$^{19}$Na - $^{19}$O} pair has an MED of \mbox{-750 keV}, due to the fact that both the $5/2^+$  ground state and the $1/2^+$
  excited state are proton unbound. The latter has a width of 110~keV,  therefore the very large spatial  extension of the 1s$_{1/2}$ proton
  wave function  should be the  soley responsible for  the huge value of the MED.   Similar arguments apply to the 
  \mbox{$^{14}$O - $^{14}$C}   (MED= -669 keV)  \cite{a14} and \mbox{$^{12}$O - $^{12}$Be} (MED= -630 keV) \cite{a12} mirror pairs.

   The calculated B(E2; 2$^+ \rightarrow 0^+_{gs}$) for $^{36}$Ca is very small, 4.7~e$^2$fm$^4$ (the Dufour-Zuker \cite{Dufour:1996hz}
  effective charges e$_{\pi}$~=~1.31e and e$_{\nu}$~=~0.46e have been used).  
  In fact this value is the smallest of all the Calcium isotopes together with that of $^{50}$Ca, 7.5$\pm$0.2~e$^2$fm$^4$~\cite{valiente}.  
  The 2$^+$ decay to the intruder 0$^+$ is suppressed by a factor  
  2$\times$10$^4$ with respect to the decay to the ground state due to the phase space factor.
  For completeness, our prediction for the  \mbox{B(E2; $2^+ \rightarrow 0_{gs}^+$)}  in $^{36}$S is 19.5~e$^2$fm$^4$, 
  which is in good agreement with the experimental value, 17.7$^{+1.7}_{-1.0}$~$e^2fm^4$~\cite{pritychenko}.  
  The calculated $\rho^2(E0)$ for  the decay of the $0^+_2$ state to the ground state  in $^{36}$Ca is $40 \times 10^{-3}$, which corresponds to a lifetime of $\tau(E0)=8.3$~ns. An effective isoscalar $E0$  charge of 1.0~e has been assumed. This effective $E0$ charge has been deduced from the known experimental value  in $^{36}$S, where the 0$^+_2$ has been observed to decay directly via an $E0$ transition to the 0$^+$ ground state and its half-life has been measured to be 8.8~$\pm$~0.2~ns~\cite{olness}, no $\gamma$ transition has been observed from the 0$^+_2$ state to the 2$^+$ state. Therefore, using eq. 1 of Ref.~\cite{Schwerdtfeger:2009fz}, we can compute an upper limit for the $\rho^2(E0)$, considering an experimental sensitivity limit of 1\% for the  0$^+_2$  to 2$^+$ decay branch  $\rho^2(E0) = \frac{I(E0)}{I_{\gamma}(E2)}\times\frac{1}{\Omega(E0)}\times\frac{1}{\tau_{\gamma}} = 9\times10^{-3}$. The electronic  $\Omega(E0)=8.7\times10^{9} s^{-1}$ factor has been calculated with BrIcc~\cite{bricc}.

As  discussed previously, the $T=2$ mirrors $^{36}$Ca and $^{36}$S are known experimentally very unequally. While the $^{36}$Ca isotope is the heaviest acknowledged $T_z = -2$ nucleus, just two neutrons away from the proton drip line, the $^{36}$S is stable. The excitation energy of the 2$^+$ state, for the $N=20$ $^{36}$Ca isotope, was measured both at GSI~\cite{Doornenbal:2007cv} and GANIL~\cite{Burger:2012gh} by using a knock-out reaction from a secondary $^{37}$Ca beam. In these two experiments a unique $\gamma$ was observed at an energy of 3015~(16)~keV at GSI and 3036~(11)~keV at GANIL.  
The momentum distribution measurement of  the $^{36}$Ca at GANIL with the SPEG spectrometer, indicates a $\ell = 0,  \ell = 2$ character of the excited 2$^+$ state, that agrees well with our calculations, where the 2$^+$ state has a dominant $sd$ configuration. 
This is all the information that currently exists for the $T_z = -2$ $^{36}$Ca, in contrast, the experimental information available for the stable $^{36}$S is copious. Over the last 
decades many reactions have been used to study the semi-magic nature of this ($N=20$) isotone.

The intruder 0$^+_2$ state in the mirror $^{36}$S, that has mainly a neutron $(sd)^{10}$-$(pf)^2$ nature,  was selectively populated via a two-neutron transfer reaction $^{34}$S(t,p)$^{36}$S~\cite{olness}, $^{34}$S(t,p$\gamma$)$^{36}$S~\cite{samworth}, as well as via a less selective reactions such as inelastic scattering with protons and $\alpha$ particles: $^{36}$S(p,p)$^{36}$S and $^{36}$S($\alpha$,$\alpha$)$^{36}$S~\cite{hogenbirk}. 
Considering the proton nature, $(sd)^{10}$-$(pf)^2$ predicted by our calculations, of the intruder 0$^+_2$ state in $^{36}$Ca, one could experimentally access this state by using a two-proton transfer reaction with a radioactive $^{34}$Ar beam, such as $^{34}$Ar($^{3}$He,n)$^{36}$Ca. The $0^{+}_{2}$ would decay directly to the $0^+_{gs}$ with a 2.63 MeV $E0$ transition with an expected lifetime of 8.3~ns.  The internal pair formation, according to  BrIcc~\cite{bricc} calculations is more than 50 times larger than the internal conversion.  
 For the T=4, T$_z=+4$, $^{32}$Mg which represents the pivotal nucleus in the $N=20$ IoI, the second 0$^+$ state was also populated via a two-neutron transfer reaction since it presents a neutron nature~\cite{wimmer}. While, the second 0$^+$ state in the T=3, T$_z=+3$, $^{34}$Si was directly observed via $\beta$ decay of a 1$^+$ isomer in $^{34}$Al~\cite{rotaru}, so in this case no transfer reaction was needed to measure the properties of the intruder state. For the $^{36}$Ca isotope, one cannot populate the intruder state via $\beta$ decay since its progenitor $^{36}$Sc is unbound. 

\begin{table}[h]
\caption{\label{tab:A34} Theoretical excitation energies (in MeV) and MED's (in keV). In the column labeled "$A=34$ $T=3$" we list the results of a 
 calculation without the Coulomb interaction.}
\begin{tabular*}{\linewidth}{@{\extracolsep{\fill}}|ccccc|}
\hline  
J$^{\pi}$ & A=34 T=3 &$^{34}$Ca &  $^{34}$Si  & MED \\ 
\hline
 $0^+_1$  & 0.0   &  0.0     &  0.0  &            \\ 
 $0^+_2$  &  2.57 &  2.33  &  2.75 &     -420 \\ 
 $2^+_1$  & 3.45  &  3.20  &  3.62 &     -420  \\ 
  $2^+_2$ &  4.46 &  4.43  &  4.49 &     -60 \\ 
\hline  
\end{tabular*}
\end{table}

  As a purely academic exercise, because of their almost certain unbound nature, we examine now  what happens when we 
  go to the  $A=34$ $T=3$ and $A=32$ $T=4$  mirrors, where the intruder configurations become more significant.
  The theoretical results for the $A=34$ $T=3$ mirrors $^{34}$Ca and  $^{34}$Si  are displayed in  Table~\ref{tab:A34}. The ground state 
  and the  $2^+_2$ are dominated by  the "normal"  $sd$ configurations   0d$_{5/2}^6$  and  0d$_{5/2}^5 $1s$_{1/2}^{1}$
  respectively.
  This results in the small   MED of the  $2^+_2$. The intruders  $0^+_2$  and $2^+_1$   decreases by 400~keV in $^{34}$Ca
  with respect to the no-Coulomb result, as they did in $^{36}$Ca. However, at difference with what happened in  $^{36}$S, 
  this does not add  constructively with a large pure sd-shell effect in  $^{34}$Si  and the  
  resulting MED's are very large but not huge.        
  
 \begin{table}[h]
 \caption{\label{tab:A32} Theoretical  excitation energies (in MeV) and MED's (in keV). In the column  labeled ''$A=32$ $T=4$'' we list the results of a 
 calculation without the Coulomb interaction.  In the right  side box,  we give the  amplitudes of the np-nh  configurations (in percentage)
  for the calculation without the Coulomb contribution.}
\begin{tabular*}{\linewidth}{@{\extracolsep{\fill}}|ccccc|ccc|}
\hline  
J$^{\pi}$ & A=32 T=4 &$^{32}$Ca &  $^{32}$Mg  & MED & 0p-0h & 2p-2h &   4p-4h\\ 
\hline
 $0^+_1$  &  0.0    &  0.0    &  0.0  &   & 9 & 54 &35    \\ 
 $2^+_1$  &  0.85  & 0.77   &  0.85 & -80 &  2 &46 &50    \\ 
 $0^+_2$  &  1.20  &  1.18  &  1.20  &  -20  &  33  &12 & 54  \\ 
 $0^+_3$ &  1.91  &   2.09  &  1.91  &  180  & 48 & 37 & 15    \\ 
\hline  
\end{tabular*}
\end{table}    
 
\begin{figure}
\begin{center}
\includegraphics[width=1.2\columnwidth,angle=0]{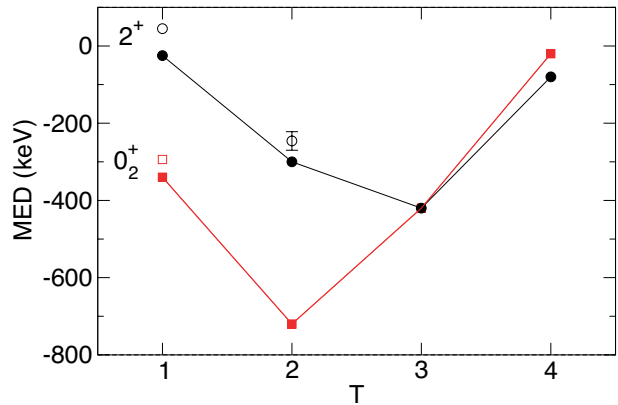}
\end{center}
\caption{(Colour online) The MED (in keV) as a function of the isospin of the mirror pair for the proton rich calcium isotopes. Open and full symbols indicate experimental data and theoretical predictions, respectively. For the T=1 mirror pair the experimental errors are within the symbols.\label{med}}
\end{figure}

 Even farther beyond the proton-drip line would eventually sit $^{32}$Ca, the mirror of the prominent member of the  $N=20$ IoI 
 $^{32}$Mg.   In Table~\ref{tab:A32} we give  the (rather exotic) structure of the low-lying states according 
 to the no-Coulomb calculation. The  only state  dominated by the normal (closed $N=20$ or closed $Z=20$) configurations 
 is the   $0^+_3$.  Due to the presence of 4p-4h configurations in addition to the 2p-2h ones, the two lowest states have 
 quite small MED's. Only the  $0^+_3$ has a large  MED due to  its mainly spherical nature. In fact, when we include the Coulomb interaction
 in the calculation,  the percentage of  
 the 0p-0h configuration  in  the $0^+_3$ of $^{32}$Ca increases to 70\%.  For this state only, the Coulomb interaction induces important differences in
 the  structure of the wave functions of the  two mirrors, due to the quasi degeneracy of the different np-nh configurations before their mixing
 by the nuclear interaction.
  The evolution of the MED's as a function of the isospin of the mirror pair $T$ is shown in Fig.~\ref{med}. 
 All in all, it seems that the CMED is elusive and the opportunity to observe it
 might be confined to the $A=36$ $T=2$ mirrors.

In summary, we predict a first excited 0$^+_2$ state at 2.7 MeV in $^{36}$Ca, 250 keV below the first 2$^+$,  in the framework of Shell Model with Configuration Interaction, using the effective interaction sdpfu-mix. This large decrease in the excitation energy of the intruder 0$^+_2$ state gives origin to the largest ever predicted Mirror Energy Difference between bound states; -720~keV, that we name Colossal MED (CMED). The calculated  B(E2; 2$^+ \rightarrow 0^+_{gs}$) transition probability of 4.7~e$^2$fm$^4$,  represents the smallest value in the calcium isotopic chain.  The theoretical $\rho^2(E0)=40 \times 10^{-3}$, leads to a  lifetime  $\tau(E0)=8.3$~ns for the intruder 0$^+$ state. According to our calculations, 
disregarding  energy threshold considerations, the CMED would  not be present in the more exotic mirror pairs   $A=34$ $T=3$ and $A=32$ $T=4$.
 A two-proton transfer reaction, such as $^{34}$Ar($^{3}$He,n)$^{36}$Ca, will give access to the 0$^+_2$ intruder state that is predicted to have a proton $(sd)^{10}$-$(pf)^2$ configuration.

{\bf Acknowledgements.} Partially supported by  MINECO (Spain) grants FPA2014-57196, FPA2015-71690-P and FPA2014-57196-C5, the Severo Ochoa Programme SEV-2016-0597 and SEV-2014-0398 and by the Generalitat Valenciana PROMETEO II/2014/019 and E.C. FEDER funds.

\bibliography{36Ca}

\end{document}